\documentclass[journal]{IEEEtran}
\usepackage{graphicx}
\usepackage{dcolumn}
\usepackage{bm}
\usepackage{amssymb}
\usepackage{amsmath}
\usepackage{hyperref}
\usepackage{cite}
\usepackage{tabularx}
\usepackage{tikz}
\usepackage{svg}
\usepackage{caption}
\usepackage{subcaption}

\usepackage{textcomp}

\usepackage[switch]{lineno} 

\usepackage{setspace}

\usepackage[pscoord]{eso-pic}

\newcommand{\placetextbox}[3]{
  \setbox0=\hbox{#3}
  \AddToShipoutPictureFG*{
    \put(\LenToUnit{#1\paperwidth},\LenToUnit{#2\paperheight}){\vtop{{\null}\makebox[0pt][c]{#3}}}%
  }%
}%

\begin{document}
\title{Development of a CsI Calorimeter for the Compton-Pair (ComPair) Balloon-Borne Gamma-Ray Telescope}

\author{Daniel~Shy,
        Richard~S.~Woolf,
        Clio~C.~Sleator,
        Eric~A.~Wulf,
        Mary~Johnson-Rambert,
        Emily~Kong,
        J.~Mitch~Davis,
        Thomas~J.~Caligiure,
        J.~Eric~Grove,
        and Bernard~F.~Phlips
        \thanks{This work was sponsored by NASA-APRA (NNH14ZDA001N-APRA, NNH15ZDA001N-APRA, and NNH1870A001N-APRA). The ComPair balloon flight is supported under NNH21ZDA001N-APRAD. Shy is supported by U.S. Naval Research Laboratory's Karles Fellowship.}
        \thanks{D. Shy, R. Woolf, C. Sleator, E. Wulf, M. Johnson-Rambert, J. M. Davis, J. E. Grove, B. Phlips are with the Space Science Division, U.S. Naval Research Laboratory, 4555 Overlook Ave., SW, Washington, DC, 20375, United States of America}
                \thanks{E. Kong is with the Technology Service Corporation, Arlington, VA,
22202, United States of America}

        \thanks{T. Caligure is with the Naval Research Enterprise Internship
Program (NREIP) at the U.S. Naval Research Laboratory}

}
\maketitle
\pagestyle{empty}
\thispagestyle{empty}



\begin{abstract}

There is a growing interest in astrophysics to fill in the observational gamma-ray MeV gap. We, therefore, developed a CsI:Tl calorimeter prototype as a subsystem to a balloon-based Compton and Pair-production telescope known as ComPair. ComPair is a technology demonstrator for a gamma-ray telescope in the MeV range that is comprised of 4 subsystems: the double-sided silicon detector, virtual Frisch grid CdZnTe, CsI calorimeter, and a plastic-based anti-coincidence detector. The prototype CsI calorimeter is composed of thirty CsI logs, each with a geometry of $1.67 \times 1.67 \times 10 \ \mathrm{cm^3}$. The logs are arranged in a hodoscopic fashion with 6 in a row that alternate directions in each layer. Each log has a resolution of around $8 \%$ full-width-at-half-maximum (FWHM) at $662 \ \mathrm{keV}$ with a dynamic energy range of around $250\ \mathrm{keV}-30 \ \mathrm{MeV}$. A $2\times2$ array of SensL J-series SiPMs read out each end of the log to estimate the depth of interaction and energy deposition with signals read out with an IDEAS ROSSPAD. We also utilize an Arduino to synchronize with the other ComPair subsystems that comprise the full telescope. This work presents the development and performance of the calorimeter, its testing in thermal and vacuum conditions, and results from irradiation by $2-25 \ \mathrm{MeV}$ monoenergetic gamma-ray beams. The CsI calorimeter will fly onboard ComPair as a balloon experiment in the summer of 2023.

\end{abstract}

\section{Introduction}
\label{sec:intro}
\placetextbox{0.5}{0.05}{\large\textsf{Distribution Statement A: Approved for public release. Distribution is unlimited.}}%

\IEEEPARstart{T}{here} are several space-based gamma-ray telescopes in the concept and development phase to address the MeV gap in astronomical observations. These efforts include the Compton Spectrometer and Imager (COSI)~\cite{COSI}, Galactic Explorer with a Coded aperture mask Compton telescope (GECCO)~\cite{GECCO}, e-ASTROGAM~\cite{astrogam}, Advanced Particle-astrophysics Telescope (APT)~\cite{APT}, SMILE-2+~\cite{SMILE}, and the All-sky Medium Energy Gamma-ray Observatory (AMEGO)~\cite{amego}, which is now adapted to AMEGO-X~\cite{amegox}. This work presents the development of a thallium-doped cesium iodide (CsI:Tl) based calorimeter in support of the ComPair balloon-based telescope~\cite{compair}, which serves as a prototype to AMEGO and will be flown in the summer of 2023. The AMEGO concept consists of four subsystems: the double-sided Silicon strip detectors (DSSD)~\cite{DSSD}, virtual Frisch-grid CdZnTe calorimeter~\cite{CZT}, CsI calorimeter, and a plastic-based anti-coincidence detector (ACD)~\cite{compair}. The main concept behind this `hybrid' telescope design is to maintain high detection sensitivity and good imaging performance through a large energy range that includes the Compton and pair-creation region. The tracker layer has the dual objective of measuring the energy of a gamma-ray Compton scatter as well as tracking electromagnetic showers during a pair-creation event. Next, the CZT, with its high resolution in the region below 10 MeV could capture the scattered photon. Higher energy events may pass through the CZT to get detected by the CsI calorimeter. The concept, therefore, has a dual use of both Compton and Pair imaging combining the COSI/COMPTEL region and that of the \textit{Fermi} Large Area Telescope.

\begin{figure}[h!]
  \centering
  \includegraphics[trim={0cm 0cm 0cm 0cm}, clip, width=\linewidth]{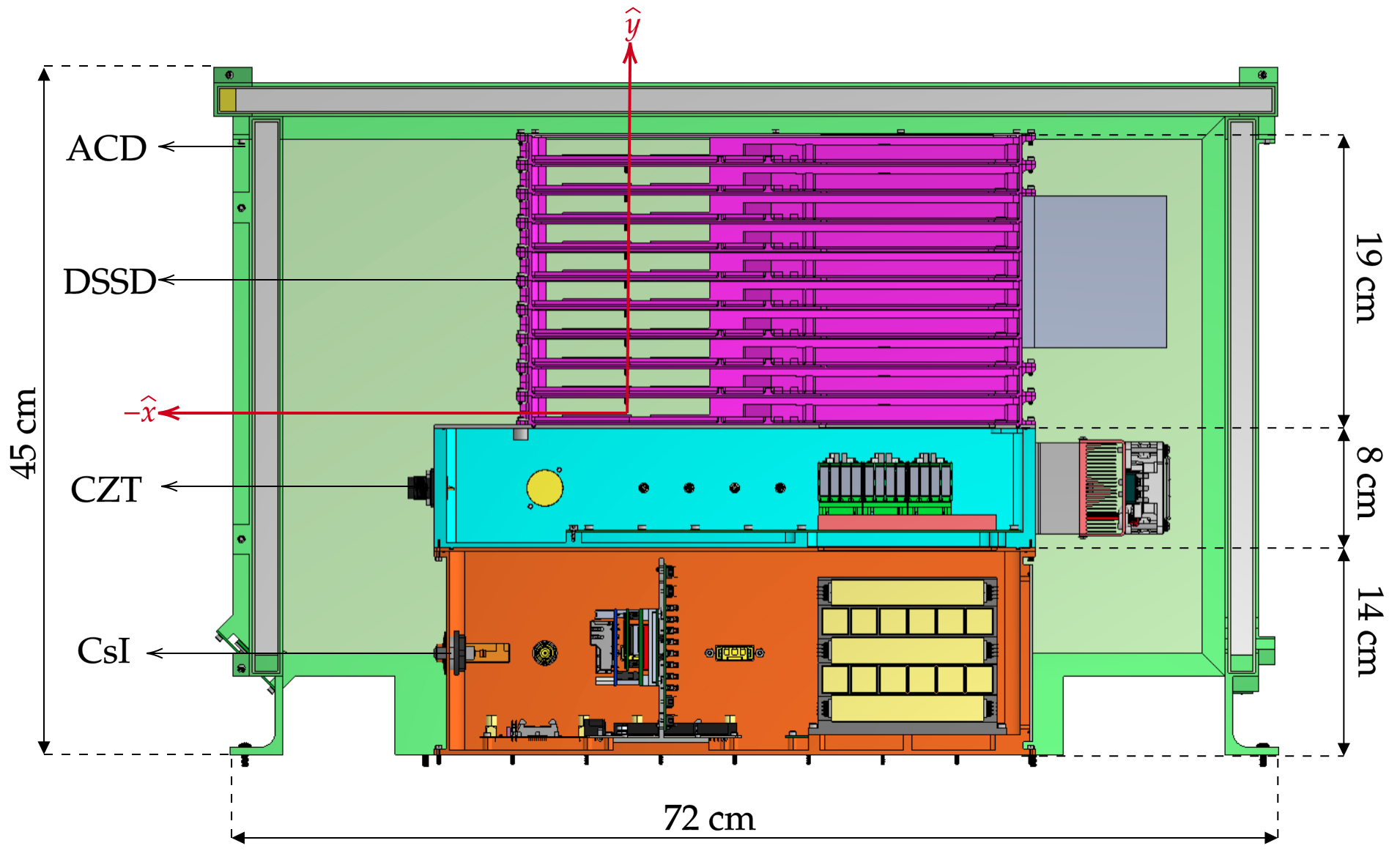}
  \caption{Cutaway view of the CAD showing ComPair’s subsystems and their dimensions. Figure is adopted from~\cite{compair}.}
  \label{fig:comPair}
\end{figure}

Fig.~\ref{fig:comPair} presents a computer aided design (CAD) model of the ComPair balloon instrument with the different subsystems labeled. The CsI calorimeter box is at the bottom of the stack colored in orange. The ComPair CsI calorimeter inherits its design and concept from the calorimeter on board the \textit{Fermi} Large Area Telescope (LAT)~\cite{GlastCsI, GLASTCalorimeter, glast}. The major difference between the LAT calorimeter and that of ComPair is the usage of SiPMs rather than PIN diodes. The diodes utilized in the LAT calorimeter did not have the necessary lower threshold required in ComPair and therefore explored using SiPMs.

This manuscript presents an expansion and finalization of the work presented in~\cite{woolfCsI}, which contains preliminary results during an early phase of development. Sec.~\ref{sec:hardware} presents the hardware and implementation. Sec.~\ref{sec:pipeline} presents the calorimeter calibration pipeline. Sec.~\ref{sec:performance} presents the calorimeter's performance and its characteristics. Sec.~\ref{sec:tvac} has results from a thermal-vacuum test. Sec.~\ref{sec:HIGS} presents results from $2-25 \ \mathrm{MeV}$ gamma-ray beam test.

\section{Hardware}
\label{sec:hardware}

The ComPair CsI calorimeter consists of 30 thallium doped CsI logs that are $1.67 \times 1.67 \times 10 \ \mathrm{cm}^3$ in volume that were manufactured by Saint-Gobain. Each end contains a $2\times2$ array of $6 \times 6 \ \mathrm{mm}^2$ SensL J-series SiPMs by onsemi (ARRAYJ-60035-4P-PCB~\cite{jSeriesArray}). This scheme allows for the depth of interaction (DOI) estimation using~\eqref{eq:doi} while we use~\eqref{eq:energy} to calculate the energy deposited.

\begin{equation}
\label{eq:doi}
\mathrm{DOI} = \frac{Q_{left}-Q_{right}}{Q_{left}+Q_{right}}.
\end{equation}

\begin{equation}
\label{eq:energy}
E_\mathrm{ADC} = \sqrt{Q_{left} \times Q_{right}}.
\end{equation}

\noindent
In the previous equations, $Q$ represents the SiPM signal recorded by the left and right SiPM for a given log. Since the DOI is a dimensionless quantity, a position calibration will need to be applied to resolve the position of interaction. $E_\mathrm{ADC}$ represents the energy deposited that will need to be energy corrected. The energy and position reconstruction techniques take heritage from the \textit{Fermi} LAT calorimeter.

The CsI logs are roughened on all six sides with sandpaper to produce depth-dependent light output. The logs are then wrapped with two layers of Tetratex~\cite{tetratex}. The SiPMs are then epoxied at each end of the log. The logs are then arranged hodoscopically in rows of six such that each layer alternates direction and are held inside a 3D-printed structure. Fig.~\ref{fig:comPair} shows the assembly located at the right side of the box with the logs shown in yellow.

\begin{figure}[h!]
  \centering
  \includegraphics[trim={0cm 0cm 0cm 0cm}, clip, width=\linewidth]{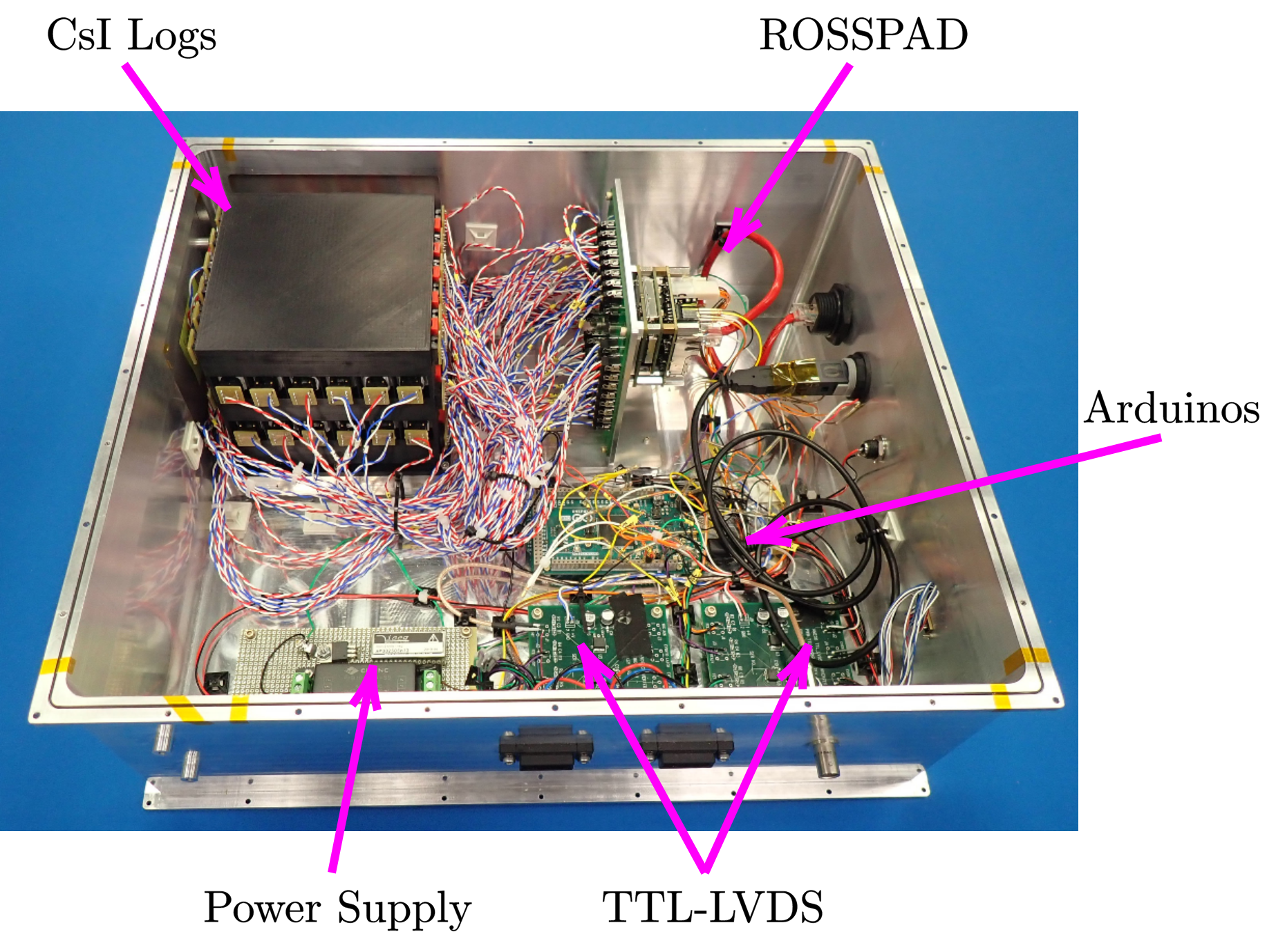}
  \caption{The CsI calorimeter system in its final configuration. Figure is adapted from~\cite{compair}.}
  \label{fig:CsICalorimeter}
\end{figure}

The SiPMs are then read out by an IDEAS ROSSPAD~\cite{rosspad}, which is a front-end that packages four SiPHRA ASICs ~\cite{SiPHRA}. We replaced the onboard power supply on the ROSSPAD with a custom unit to provide the additional current required to operate the large number of SiPMs. The SiPMs are biased to $27.5 \ \mathrm{V}$. Note that the breakdown voltage for the J-Series is between $24.2-24.7 \ \mathrm{V}$. Fig.~\ref{fig:CsICalorimeter} shows the fully assembled subsystem with the CsI assembly located in the upper left, the ROSSPAD in the upper right, and the power supply located in the lower left.

Peripheral hardware that supports the subsystem is located within the lower right quadrant. It is comprised of two Arduino Dues, which are single-board microcontrollers with a set of digital and analog input/output (I/O) pins that are commonly used in the hobbyist community~\cite{arduino}. One Arduino is utilized for housekeeping: it monitors six thermocouples that are distributed within the box as well as a real-time clock. The second Arduino is used to synchronize with the other ComPair subsystems. The synchronization between subsystems works as follows: when each subsystem detects an event, a trigger is transmitted to the ComPair trigger module~\cite{MakotoTM}. Then, based on some trigger logic (e.g., a coincidence between two or more subsystems), the module responds with a trigger acknowledgment along with a numerical Event ID. The ROSSPAD cannot receive EventIDs; instead we use a second Arduino Due  to receive trigger acknowledgments and EventIDs from the ComPair TM, in addition to receiving a trigger signal directly from the ROSSPAD. Thus, the Arduino can accurately insert the Event IDs from the trigger module into the ROSSPAD data stream. In addition to maintaining Event IDs, the CsI subsystem also takes in a GPS-disciplined pulse per second (PPS) signal which allows for the correction of the on-board clocks. The subsystem communicates to the trigger module through two TTL-LVDS boards.

The calorimeter is controlled by an external computer that records the ROSSPAD events as well as data received by the Arduino. During the balloon flight, the calorimeter will be controlled by a Versalogic BayCat single board computer~\cite{compair}.

\section{Position and Energy Calibrations}
\label{sec:pipeline}

\begin{figure}
     \centering
     \begin{subfigure}[b]{0.4\textwidth}
         \centering
         \includegraphics[width=\textwidth]{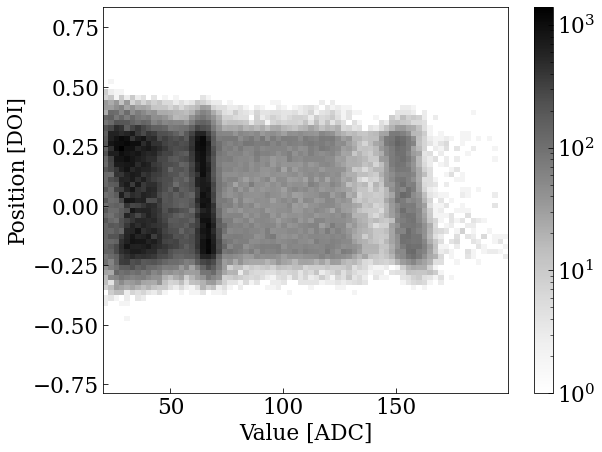}
         \caption{Raw}
         \label{fig:rawDepthEnergy}
     \end{subfigure}
     \hfill
     \begin{subfigure}[b]{0.4\textwidth}
         \centering
         \includegraphics[width=\textwidth]{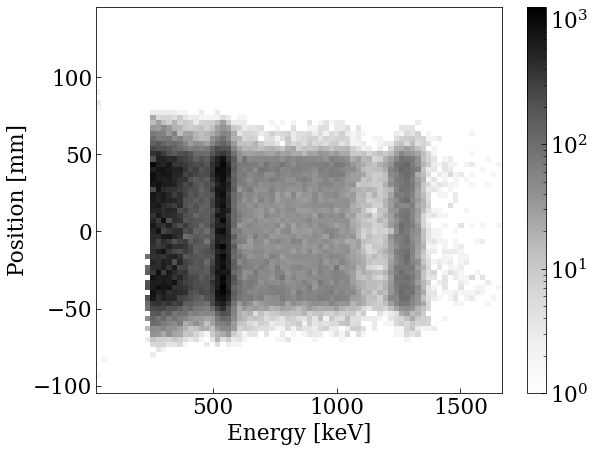}
         \caption{Corrected}
         \label{fig:correctedDepthEnergy}
     \end{subfigure}
        \caption{Depth-energy histograms of a single log's response to a $^{22}\mathrm{Na}$. (a) Presents the raw response clearly showing the $0.511$ and $1.27 \ \mathrm{MeV}$ peak and its distribution along the depth (DOI). (b) Applies the calibration routine which corrects the energy and position. The correction clearly shows the alignment of the full-energy peak across all depths due to the calibration.}
        \label{fig:depthEnergy}
\end{figure}

The position and energy calibrations utilize a multitude of variables within each log. Simply, we apply an energy correction for each DOI voxel within a log. The detailed procedure is as follows:

\begin{enumerate}
    \item Take flood irradiations with multiple sources ($^{22} \mathrm{Na}, ^{137}\mathrm{Cs}, ^{228} \mathrm{Th}$), which allows for corrections across different energies. 
    \item Develop an energy-position density plot like the one in Fig.~\ref{fig:rawDepthEnergy}. There, we locate the full-energy peak for a given DOI, i.e., the ADC value associated with the $511 \ \mathrm{keV}$ line at multiple DOIs. 
    \item Calculate the gain correction for that given DOI and repeat that for each subsequent depth and energy. For each log, we implement 150 depth bins, which equate to roughly $\sim 1 \ \mathrm{mm}$ bins. This results in a DOI-dependent calibration mask that can be applied to compensate for the different photopeak centroids. 
    \item Apply the gain calibration to convert from ADC to energy. 
    \item The conversion from the DOI to the true depth is done by bounding the DOI values for the full-energy peak and calculating the DOI's distribution centroid. This is done by creating a histogram of DOI values for a given energy, and then calculating the min/max of that distribution as well as the average. 
    \item The DOI values are then scaled and shifted to the appropriate positions and result in Fig.~\ref{fig:correctedDepthEnergy}. This scheme assumes a linear DOI response across the length of the log.
    
    \item We apply an ADC-to-energy non-linearity correction by studying the gain calibration value for each given DOI across different energies and fitting a quadratic curve across them.

\end{enumerate}

\begin{figure}[h!]
  \centering
  \includegraphics[trim={0cm 0cm 0cm 0cm}, clip, width=\linewidth]{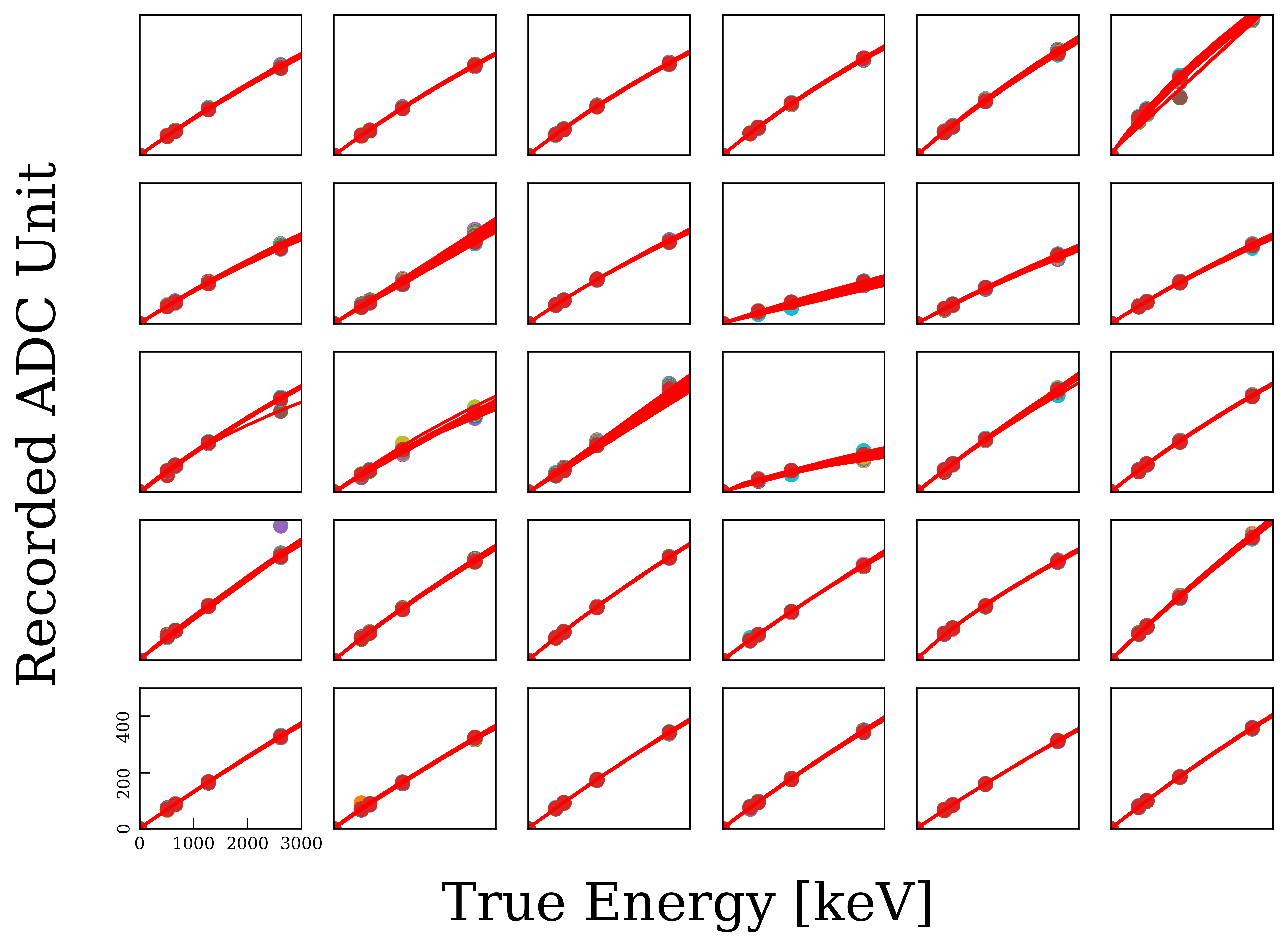}
  \caption{Quadratic calibration curves that plot the recorded ADC $(E_\mathrm{ADC})$ vs. the line energies of the radioactive lab sources for each log. The scatter points represent experimental data while the red curves represent quadratic fits. We plot the data for all depths for a given log but the traces are not very visible in most logs due to their relatively low variance such that most features are superimposed. The axis for all the subplots utilizes the same numerical range as that of the lower-leftmost subplots.}
  \label{fig:gainCalibrationFit}
\end{figure}

In step 7, the non-linearity could originate in the SiPM~\cite{siPMNonLinearity} or the readout electronics~\cite{SiPHRA}. In addition, variations in the ASIC pedestal from the calibration and actual measurement could also affect the calorimeter's response. Fig.~\ref{fig:gainCalibrationFit} plots the different calibrations for all depths and energies for each log along with the quadratic fits. As seen in the plot, most of the calibration fits do match the recorded photopeak centroid and have good agreements across depth. Finally, we also apply a correction as a function of energy on the DOI range and centroids by stretching and shifting those values across energy.

\section{Calorimeter Performance}
\label{sec:performance}

\begin{figure}[h!]
  \centering
  \includegraphics[trim={0cm 0cm 0cm 0cm}, clip, width=\linewidth]{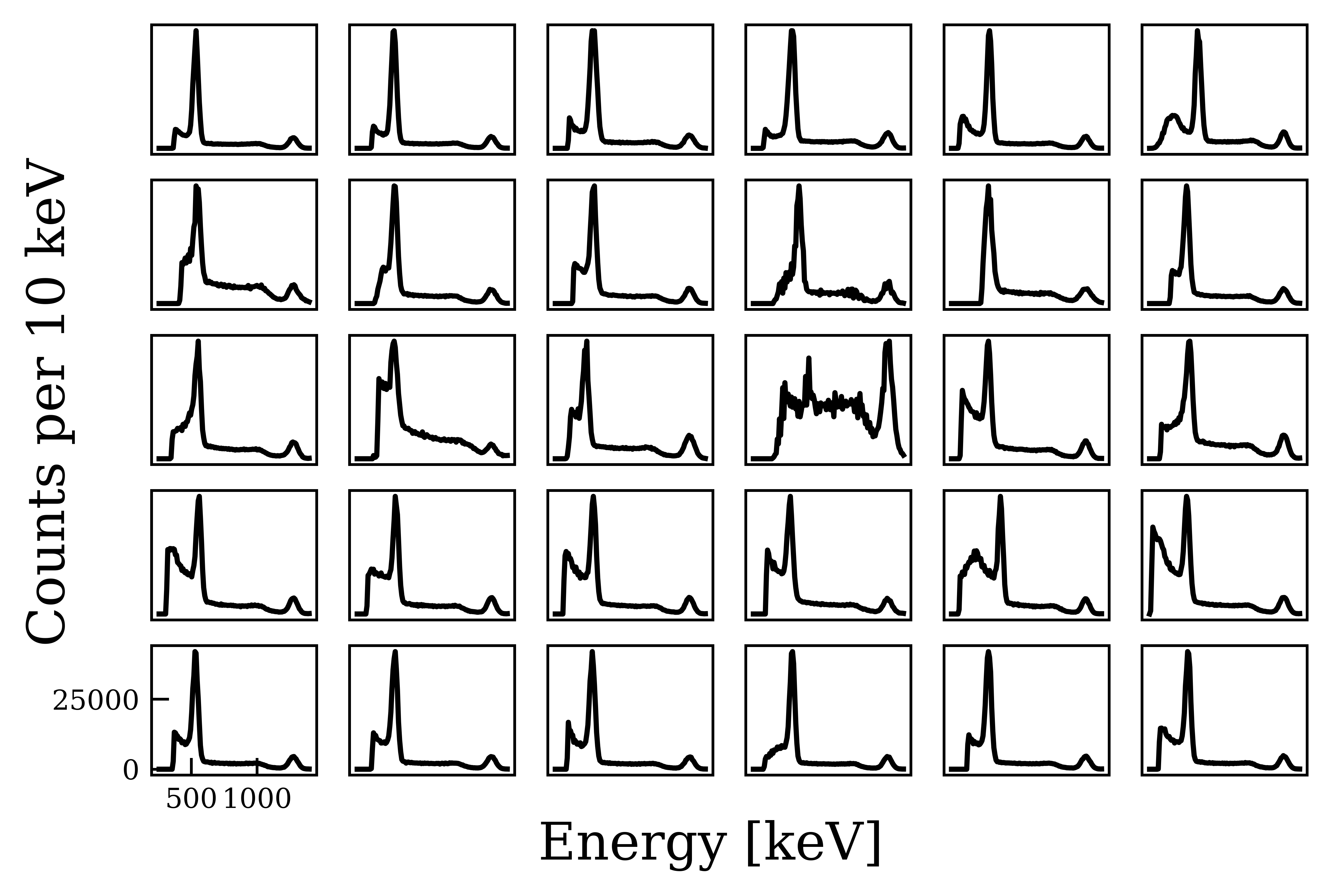}
  \caption{Corrected $^{22} \mathrm{Na}$ spectra for each log in the CsI calorimeter. Log [3,4] has a very high lower-threshold which results in a high continuum to $1.274 \ \mathrm{MeV}$ peak ratio. The source of this is still under investigation.}
  \label{fig:naSpectra}
\end{figure}

Fig.~\ref{fig:naSpectra} plots the spectra taken with a $^{22}\mathrm{Na}$ source flood irradiation for each log. The resolution averaged over all logs is $7.9\%$ FWHM at $662 \ \mathrm{keV}$ and $6.8\%$ FWHM at $1.274 \ \mathrm{MeV}$. The poor performance of several logs is still under investigation. A focused assessment revealed that their performance is not due to the SiPM or the readout electronics, but the combination of the two. Our leading theory is that there might be an issue with the connection between the SiPM and the ASIC. We note that the resolution does not scale with the inverse square root of the energy as expected; this is likely due to the accumulation of many small errors in correcting the response over multiple depths and multiple logs.

The log's depth of interaction position resolution is estimated to be around $\sim 1 \ \mathrm{cm}$ FWHM. The position resolution was estimated via a collimated fan measurement accomplished with two lead bricks separated by American (\textcent)  pennies ($1.52 \ \mathrm{mm}$). Extrapolating from the energy calibration, we estimate the upper energy range of the system to average around $30 \ \mathrm{MeV}$ using this 12-bit ADC readout~\cite{SiPHRA} with a lower energy limit of $250 \ \mathrm{keV}$. (We show the performance of the system to 25 MeV in Sec.~\ref{sec:HIGS}).

\begin{figure}
     \centering
     \begin{subfigure}[b]{0.45\textwidth}
         \centering
         \includegraphics[width=\textwidth]{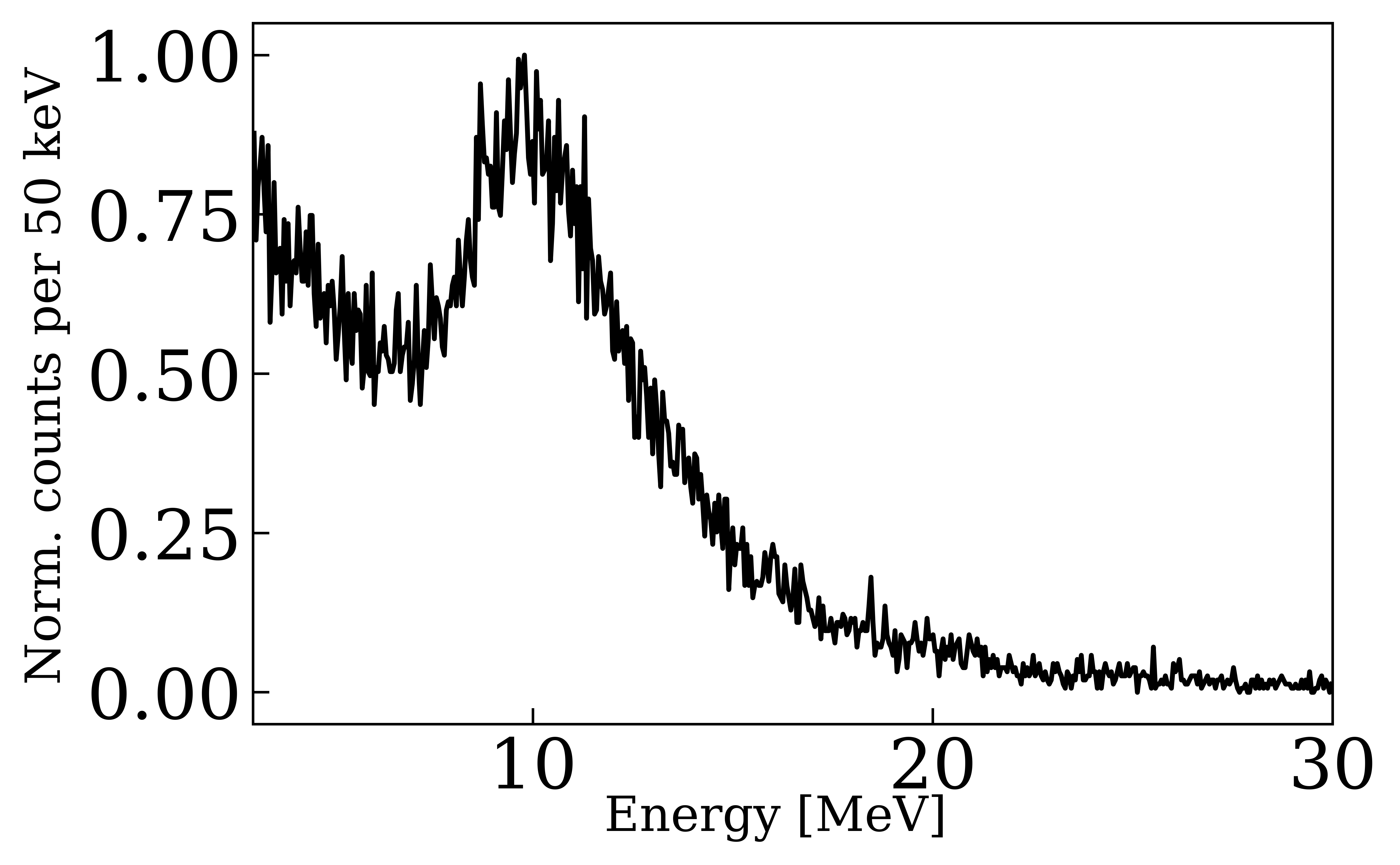}
         \caption{Single Bar Response}
         \label{fig:singleBarMuon}
     \end{subfigure}
     \hfill
     \begin{subfigure}[b]{0.45\textwidth}
         \centering
         \includegraphics[width=\textwidth]{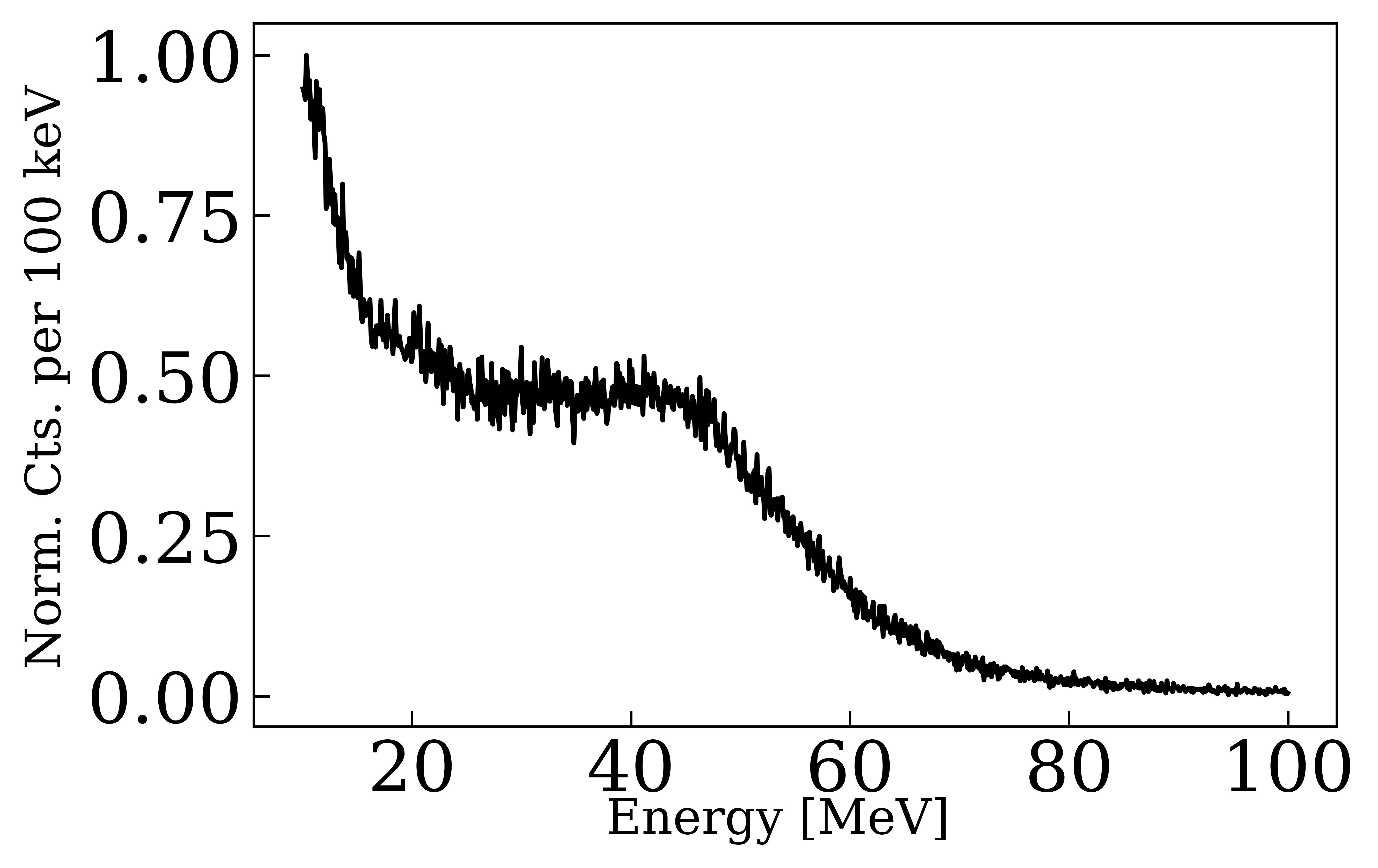}
         \caption{All Logs Summed}
         \label{fig:allBarMuon}
     \end{subfigure}
        \caption{Energy spectra of muons. (a) Presents the response of a single bar while (b) sums the interactions that occurred in all logs for a given event.}
        \label{fig:muon}
\end{figure}

We also studied the muon response of the calorimeter to demonstrate the ability of the entire calorimeter to measure events at the 10s of MeV. Fig.~\ref{fig:muon} plots the response from an overnight measurement where (a) presents a spectrum of a single log while (b) plots the spectrum of the interactions in all logs summed. In the measurement, only internal triggers were used. Both spectra present a Landau distribution and continuum with the expected energy response for muons in the calorimeter~\cite{landau}. To first order, assuming a $< -\mathrm{d}E_\mathrm{min}/\mathrm{d}x > = 1.243 \ \mathrm{MeV \ cm^2 /g}$ (the minimum in the curve ~\cite{muonStopping}), with a CsI density of $4.51 \ \mathrm{g/cm^3}$  and a thickness $1.67$, we get $9.1 \ \mathrm{MeV}$ deposition, close to the minimum in the measured distribution. Because the energy deposited by the muons is a distribution, the muon spectra will be skewed right to higher energies. 

In (b), the spectrum is roughly 5 times the energy of the distribution of (a) which is an appropriate scale since the calorimeter is 5 layers deep. Note that we did not implement any event selection criteria in (b).

\section{Environmental Testing}
\label{sec:tvac}
\begin{figure}[h!]
  \centering
  \includegraphics[trim={4cm 5cm 6cm 0cm},angle=-90,origin=c, clip, width=\linewidth]{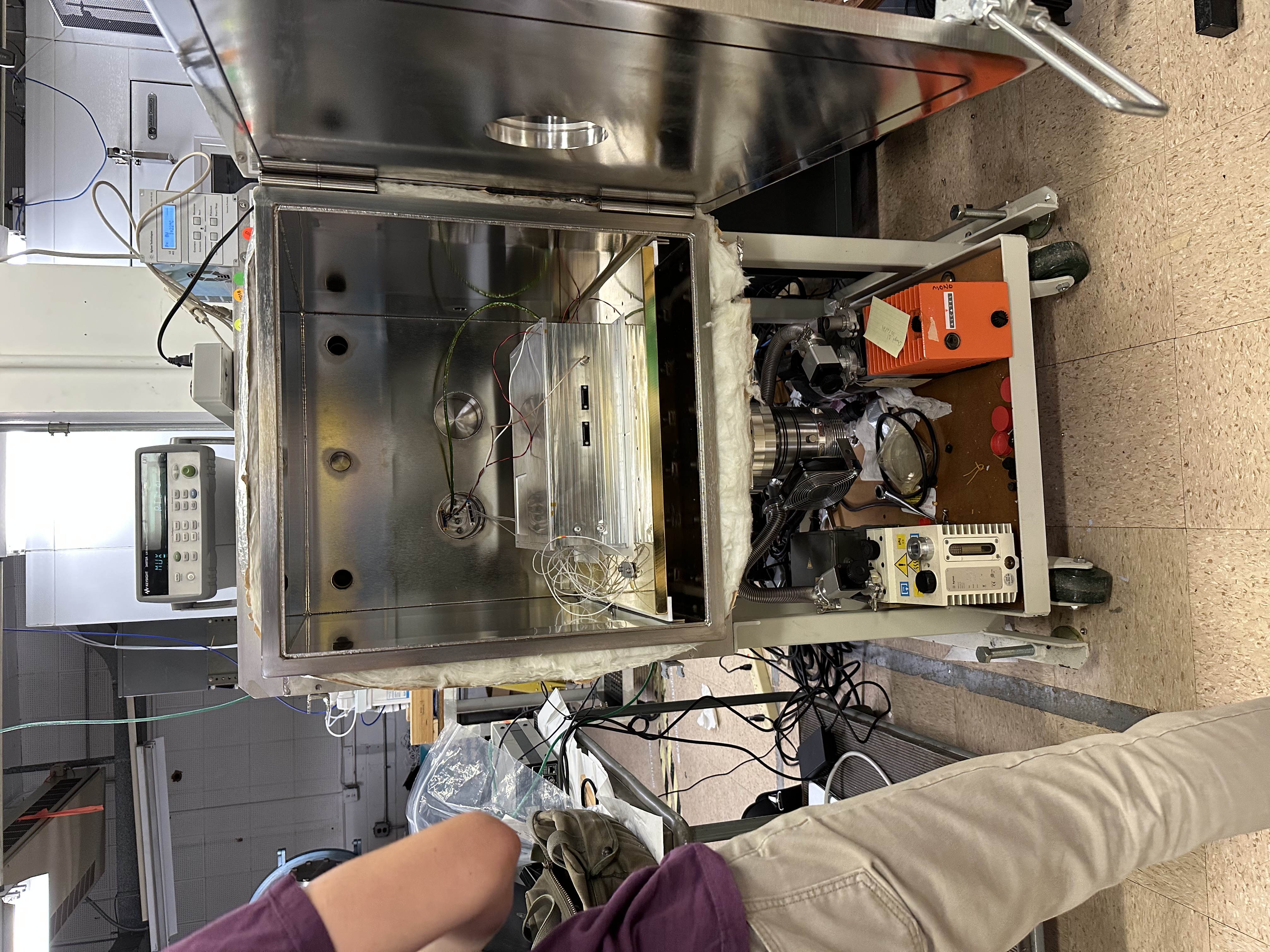}
  \caption{The CsI calorimeter located in the environmental chamber.}
  \label{fig:TVACChamber}
\end{figure}

We tested the prototype in a thermal vacuum chamber (TVAC), as is standard in most balloon-borne instruments~\cite{GLAST_Environment}. The test intends to verify the functionality and thermal design of the calorimeter instrument over the hardware temperature limits expected during a high-altitude balloon flight. The procedure was an abbreviated version of the NASA General Environmental Verification Standard  (GEVS) standard~\cite{gevs}. Fig.~\ref{fig:TVACChamber} shows the calorimeter in the chamber.

\begin{figure}[h!]
  \centering
  \includegraphics[trim={0cm 0cm 0cm 0cm}, clip, width=\linewidth]{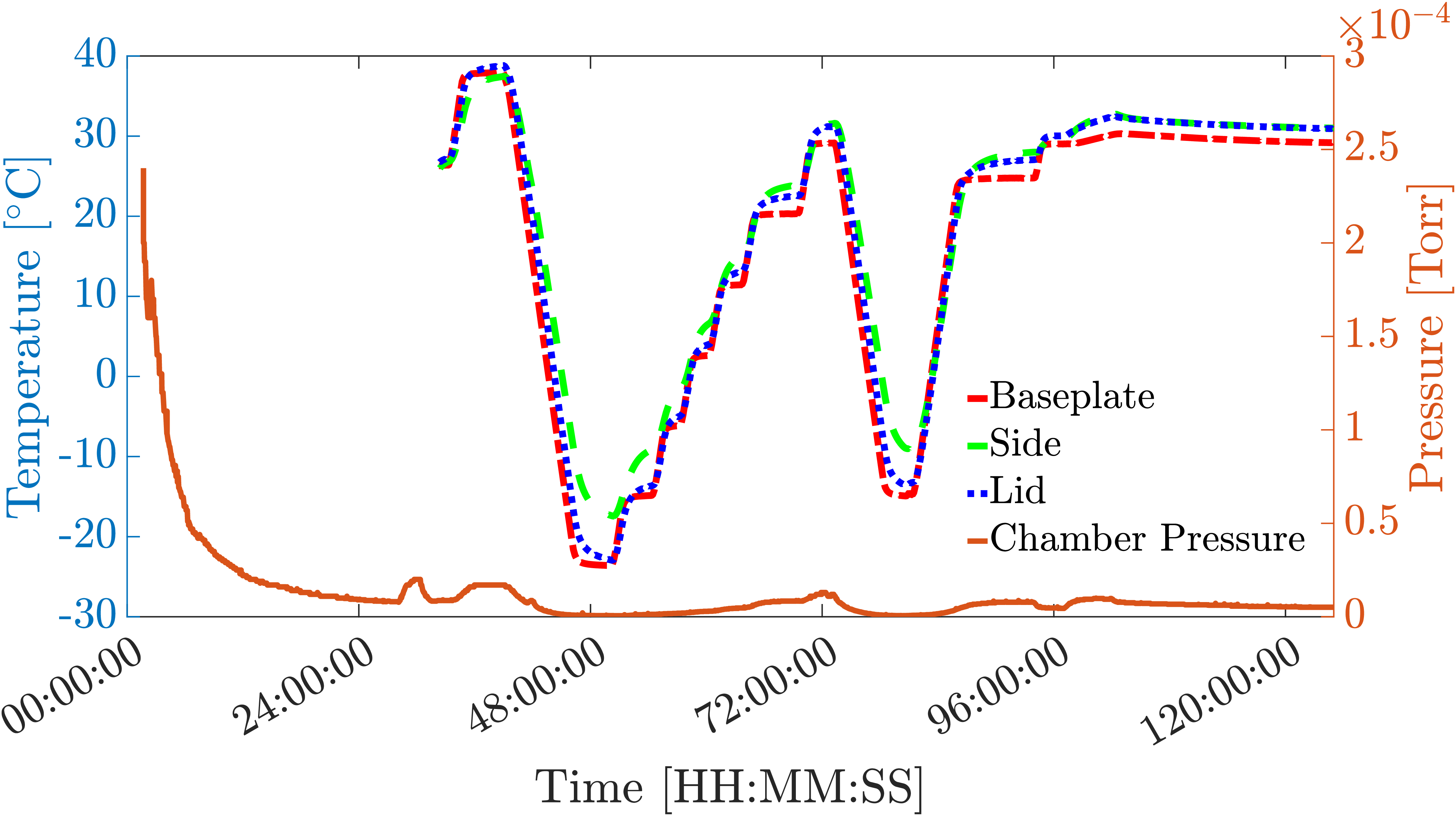}
  \caption{Environmental chamber conditions during TVAC testing. The solid line plots the chamber pressure while the dashed lines plot the temperatures experienced by the external thermocouple.}
  \label{fig:tvacChamber}
\end{figure}

Fig.~\ref{fig:tvacChamber} plots the environmental conditions that the prototype experienced during testing. The \textit{x}-axis plots the time elapsed since the vacuum pump was turned on. The solid-line trace shows the pressure in the chamber while the dashed lines show the thermocouple temperatures that are placed on the chamber's baseplate and two on the exterior of the calorimeter. We completed survival tests at $-30^{\circ}\mathrm{C}$ and $40^{\circ}\mathrm{C}$ where the system was turned off. We then conducted operational tests and characterized the system in temperatures between $-20^{\circ}\mathrm{C}$ and $30^{\circ}\mathrm{C}$. The thermal cycle includes one survival (hot and cold) cycle with one operational cycle; the plateaus lasted for four hours during the survival cycle and two hours for the operational cycles. Temperature stabilization is achieved when the control thermocouple is within $1^{\circ}\mathrm{C}$ of the soak goal temperature and does not change by more than $3^{\circ}\mathrm{C}$. The thermal portion of the TVAC lasted around 3 days.

\subsection{Temperature Response}
\begin{figure}[h!]
  \centering
  \includegraphics[trim={0cm 0cm 0cm 0cm}, clip, width=\linewidth]{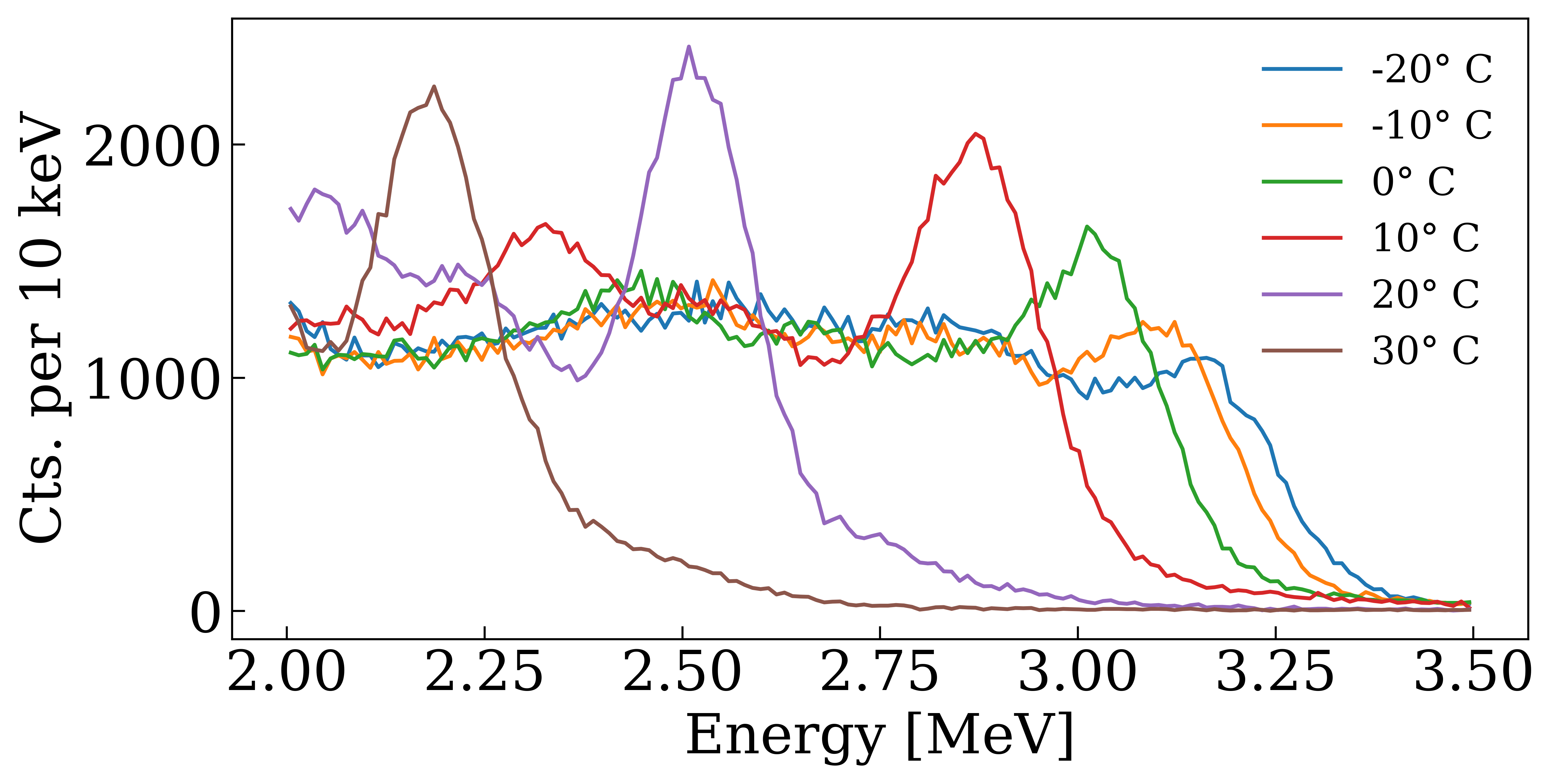}
  \caption{Thoriated-rod spectra gated at the $2.614 \ \mathrm{MeV}$ peak taken during TVAC testing. In the plot, each interaction is treated individually.}
  \label{fig:tempAllSpectra}
\end{figure}

We placed thoriated welding rods at the top and the bottom of the chamber to provide a calibration source ($2.614 \ \mathrm{MeV}$ line). The major isotope in the rods is most likely $^{232}$Th with a gamma-ray spectrum available in~\cite{thGammaSpec}. During the TVAC test, we collected a 15-minute run at several checkpoints to develop a calibration curve as a function of temperature as we may experience a variety of temperatures during the balloon flight. We observed no significant difference when the system was in one atmosphere or under a vacuum. After the hot and cold survival cycle, beginning at hour 48, we ramped up from $-20^{\circ}\mathrm{C}$ to $30^{\circ}\mathrm{C}$ in $10^{\circ}\mathrm{C}$ steps and dwelled at each of those temperatures to soak for 1.5 hours which was followed by a calibration run. Fig.~\ref{fig:tempAllSpectra} plots the energy response of the calorimeter at different temperatures where multiple interactions for a given event are not summed. The spectra used an ASIC pedestal subtraction value and energy calibration that was performed at room temperature before the start of the experiment. The trend shows that the peak centroid increases as temperature decreases which is consistent with the ASIC manufacturer's expected response. In addition, the SiPM's gain will increase with the decrease in temperature. Convolved with the response is also the crystal light output~\cite{Valentine}.

\begin{figure}[h!]
  \centering
  \includegraphics[trim={0cm 0cm 0cm 0cm}, clip, width=\linewidth]{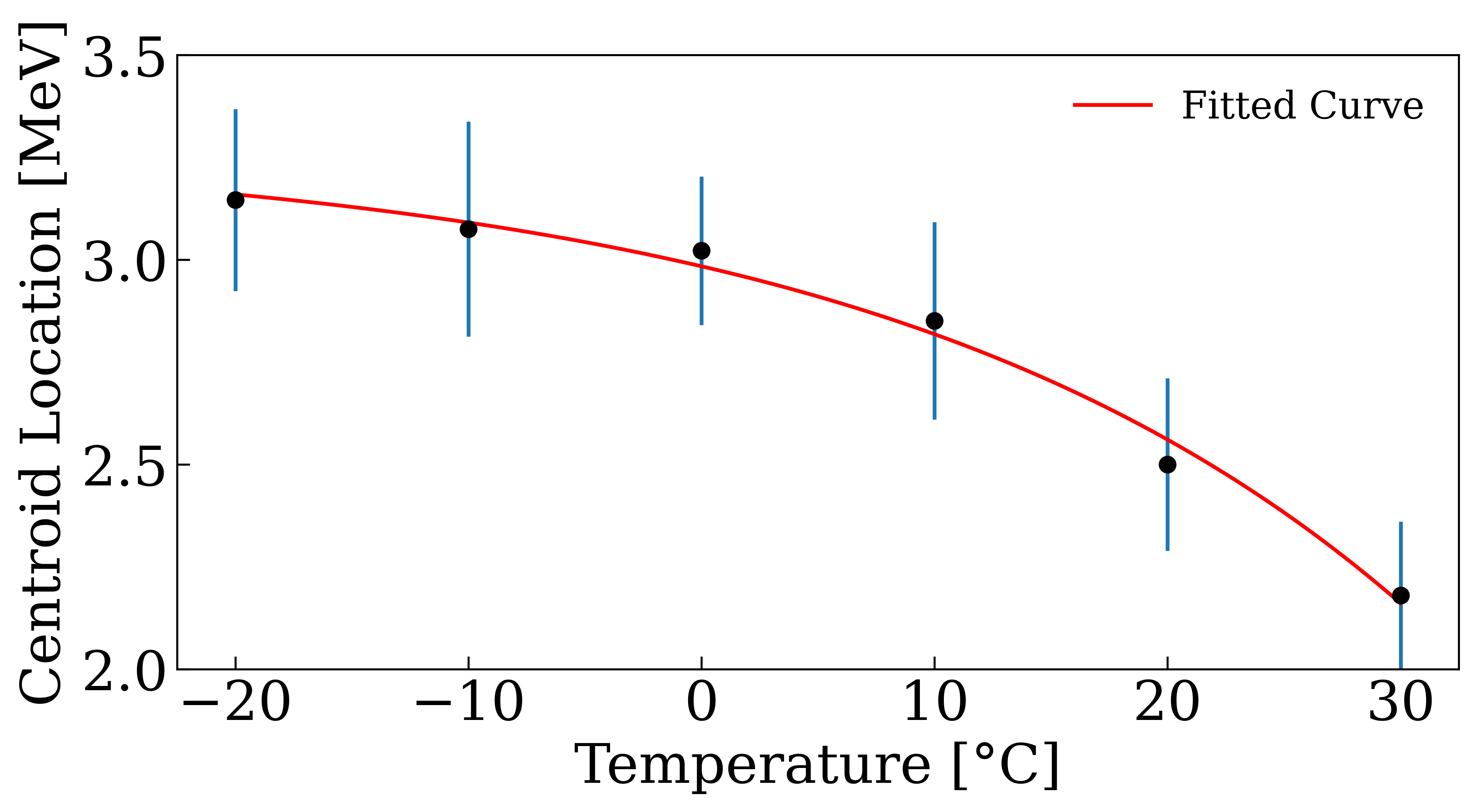}
  \caption{Calibration curve of the $2.614 \ \mathrm{MeV}$ peak as a function of temperature. The vertical bars represent the FWHM measured using the upper-half of the peak. The fit of the full-energy peak is $-0.3 e^{0.044T} + 3.28$.}
  \label{fig:calibrationCurve}
\end{figure}

Fig.~\ref{fig:calibrationCurve} plots the $2.614 \ \mathrm{MeV}$ centroid as a function of different temperatures. The vertical bars represent the FWHM calculated using only the upper half of the peak. The red line presents an exponential fit. We intend to use the fit as a gain correction across all energies for each log. We did not complete a comprehensive study for each depth-energy bin due to the limited available time we had with the TVAC chamber.

\begin{figure}[h!]
  \centering
  \includegraphics[trim={0cm 0cm 0cm 0cm}, clip, width=\linewidth]{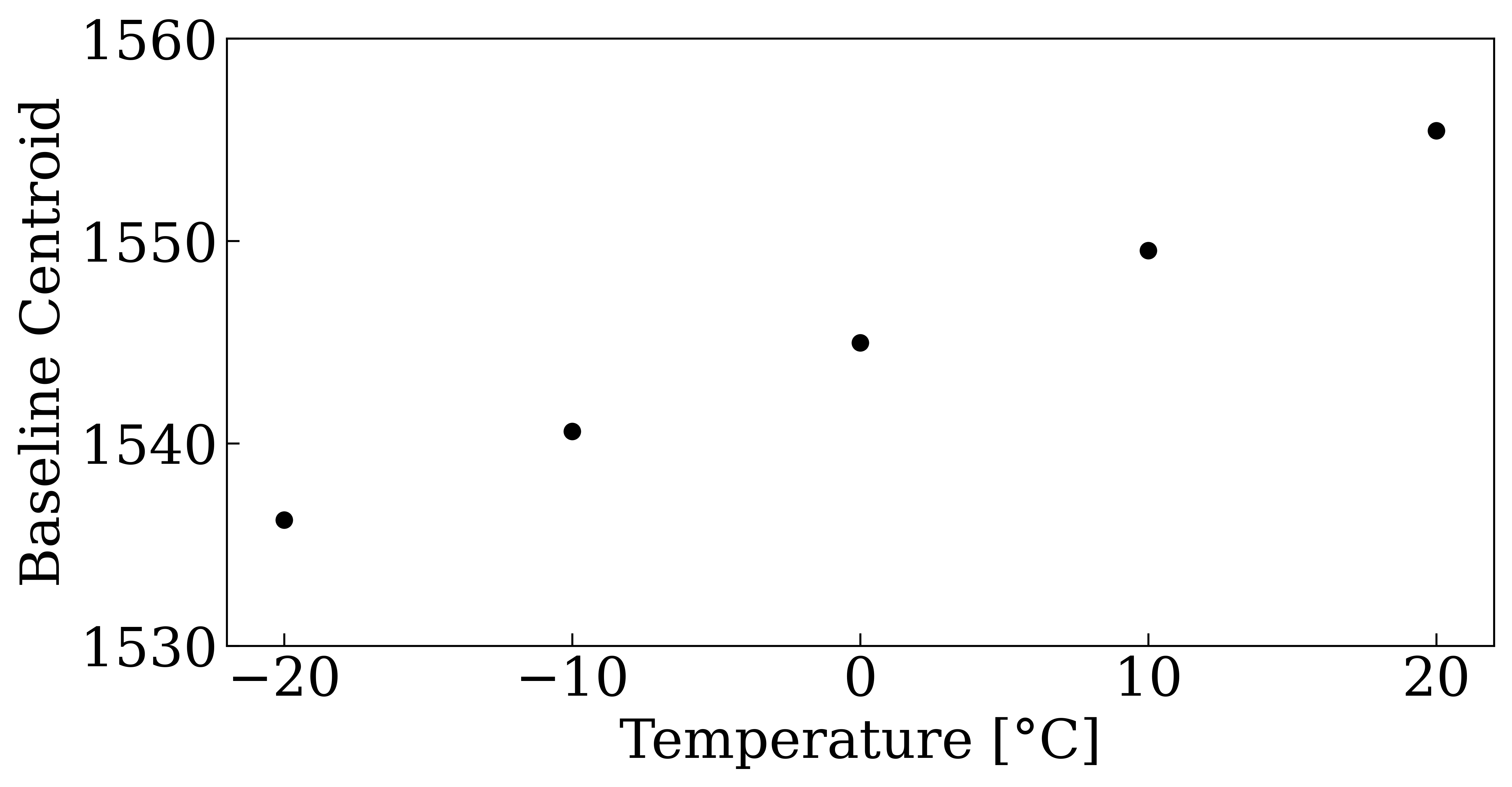}
  \caption{Baseline pedestal of the SiPHRA ASICs as a function of temperature.}
  \label{fig:baseline}
\end{figure}

Finally, the SiPHRA ASIC's baseline pedestal, which is an electrical offset that is added to each channel to prevent the value from going negative, also shows a significant temperature dependence. Fig.~\ref{fig:baseline} plots the baseline as a function of temperature and shows a linear relationship between the two. The baseline was measured by force triggering all channels and observing the baseline in each channel.

\section{Results from High-Energy Gamma-ray Beam Test}
\label{sec:HIGS}

The ComPair telescope was taken to the High Intensity Gamma-ray Source (HIGS) located at the Triangle Universities Nuclear Laboratory (TUNL). This section only presents CsI Calorimeter results while some full instrument results are presented in ~\cite{compair}. In the experiment, there are 5 layers of $60 \ \mu \mathrm{m}$ DSSDs, $2 \ \mathrm{cm}$ of CdZnTe, and additional passive material between the beam and the CsI calorimeter that equate to almost $13 \ \mathrm{g/cm}^2$ worth of grammage. This material can be seen in Fig.~\ref{fig:comPair} minus the ACD.

\begin{figure}
     \centering
     \begin{subfigure}[b]{0.4\textwidth}
         \centering
         \includegraphics[width=\textwidth]{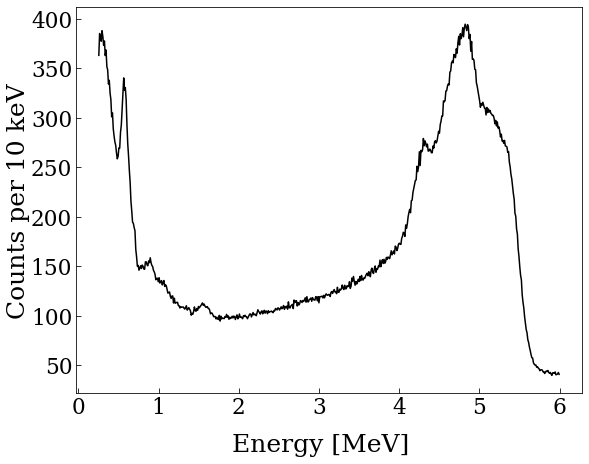}
         \caption{ }
         \label{fig:5MeVNormal}
     \end{subfigure}
     \hfill
     \begin{subfigure}[b]{0.4\textwidth}
         \centering
         \includegraphics[width=\textwidth]{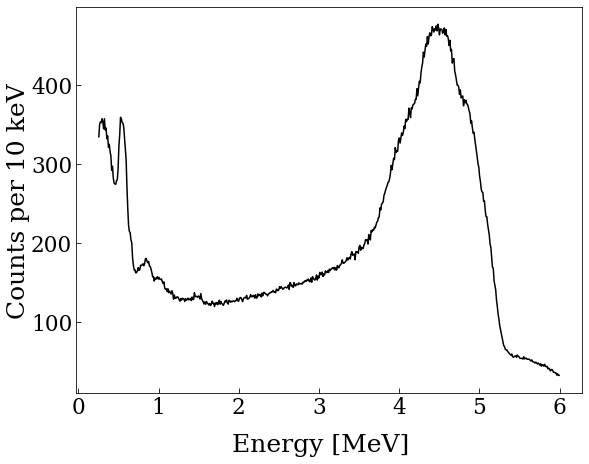}
         \caption{ }
         \label{fig:5MeV15}
     \end{subfigure}
        \caption{$5.05 \ \mathrm{MeV}$ spectra summing all the interactions. (a) Shows the spectra for a beam incident normal to instrument while (b) has the beam angles $15^{\circ}$.}
        \label{fig:5MeV}
\end{figure}

HIGS can produce mono-energetic gamma rays in a $0.5 \ \mathrm{cm}$ diameter collimated beam. The energy resolution of the beam is near $1 \%$ FWHM at $5 \ \mathrm{MeV}$. Fig.~\ref{fig:5MeV} plots the spectrum for a $5.05 \ \mathrm{MeV}$ beam with Fig.~\ref{fig:5MeV}a showing the beam incident directly normal to the calorimeter while Fig.~\ref{fig:5MeV}b has the instrument angled $15^\circ$ away from normal. The spectra show a prominent single escape peak with a hint of a double escape peak. The full-energy peak manifests as a fall-off feature. The resolution of the spectra differs in the two configurations, which is most likely due to the different number of interactions for a given event due to the detector's geometry.

\begin{figure*}[h!]
  \centering
  \includegraphics[trim={0cm 0cm 0cm 0cm}, clip, width=\linewidth]{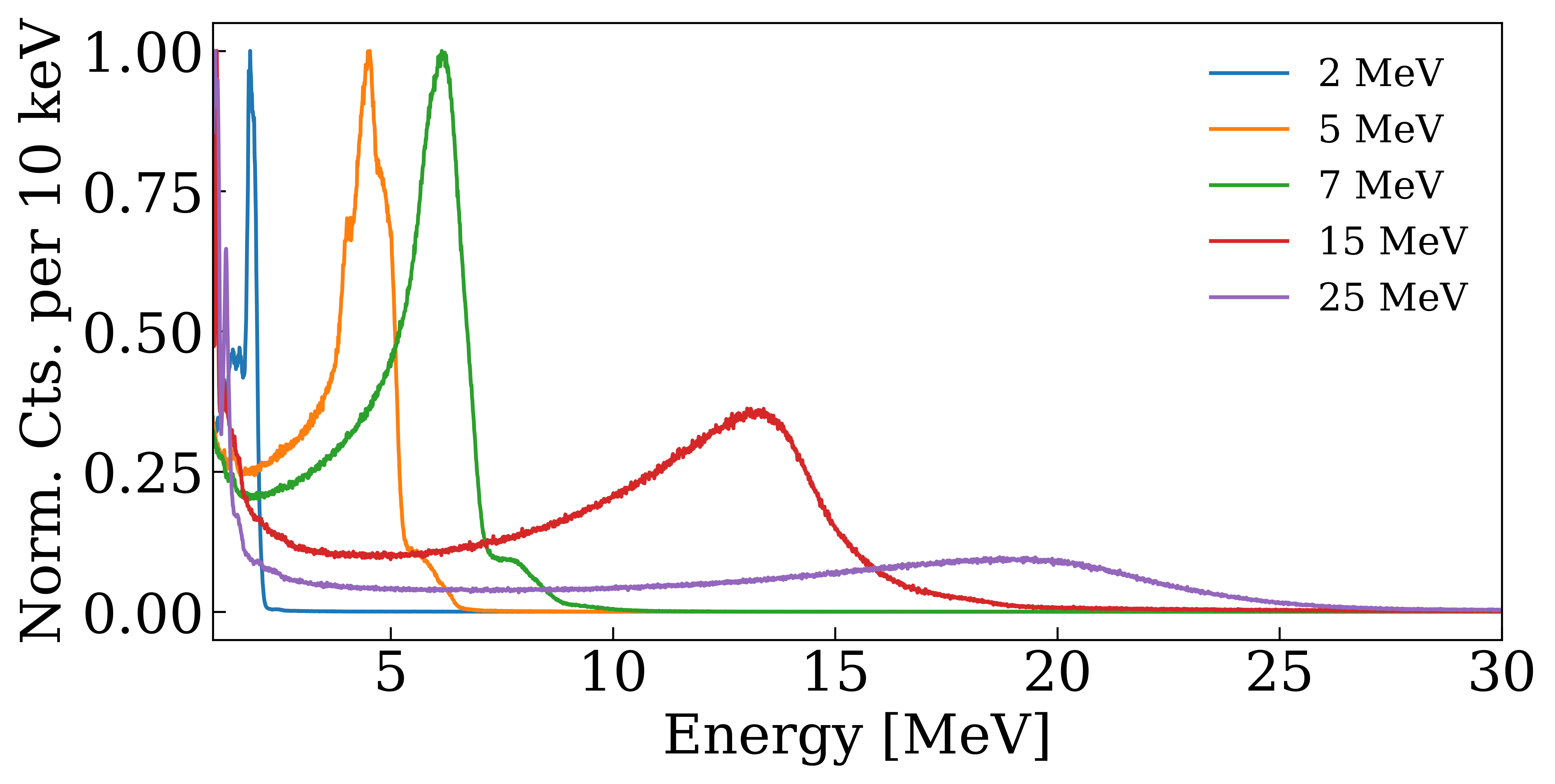}
  \caption{Spectra of the total energy deposited in the calorimeter. The beam's vector is normal to the instrument.}
  \label{fig:allHIGS}
\end{figure*}

The other energies explored include $2,7,15$ and $25 \ \mathrm{MeV}$. Fig.~\ref{fig:allHIGS} shows their spectra for a normal beam incidence. At $7 \ \mathrm{MeV}$ and above, the spectrum becomes more continuum-like and loses the full energy peak because of the underlying physics and their cross-sections where the main mode of interaction shifts from Compton to pair-production. We also observe that the full-energy efficiency decreases and the ensuing electromagnetic showers tend to escape the system. This phenomenon has been observed in other experiments~\cite{highEnergyLaBr}. Appendix~\ref{sec:sims} presents several simulations of the conducted experiment showing relativly good agreement between experiment and simulation. There, we experimentally show that the system has the dynamic range to measure gamma rays at 25 MeV.

\section{Conclusion and Future Work}
This work presents the development of a CsI-based calorimeter as part of the ComPair telescope. The instrument specifically serves as a prototype of the AMEGO/AMEGO-X mission concept. 

The calorimeter displays an energy resolution that averages around $8\%$ per log with an energy dynamic range of $0.25-30 \ \mathrm{MeV}$ extrapolated using the 12-bit IDEAS ROSSPAD. We also demonstrated the calorimeter has the energy range to measure 25 MeV gamma rays and muon traversing the entire calorimeter equating to more than 60 MeV deposition. The calorimeter, along with the rest of the ComPair telescope, is slated to fly as a short-duration high-altitude balloon experiment out of Fort Sumner, NM in the summer of 2023. The flight will be hosted by NASA’s Columbia Scientific Balloon Facility.

We are currently developing the next iteration of the prototype which is meant to resemble an AMEGO-X configuration that will make up a single tower. This includes utilizing larger logs that are slated to be $388 \ \mathrm{mm}$ along with a $15 \times 15 \ \mathrm{mm}$ cross-section. The new prototype will utilize a custom front-end readout developed specifically for the CsI calorimeter.

The SiPMs in this work did not have the desired energy range of $250 \ \mathrm{keV}-500 \ \mathrm{MeV}$. Therefore, we are developing custom SiPM readouts to address energy non-linearity and extend the dynamic range when reading out the scintillators. The SiPM readouts will contain two species of SiPMs that are optimized for either low or high energy ranges in terms of their microcell and overall size. This allows for a ``dual-gain'' approach that was inspired by the \textit{Fermi}-LAT's ``dual-gain'' PIN diodes~\cite{GLASTCalorimeter}. The development of dual-gain SiPMs will be reported on in a future publication.

\section*{Acknowledgment}
We thank Iker Liceaga-Indart from NASA Goddard Space Flight Center for developing the calorimeter housing.

We are grateful to the accelerator physics group led by Prof. Y.K. Wu at High Intensity Gamma-ray Source (HIGS), Triangle Universities Nuclear Laboratory for their aid in an excellent $\gamma$-ray beam test.

\appendix

\section{Simulation of the System}
\label{sec:sims}
\begin{figure}[h!]
  \centering
  \includegraphics[trim={0cm 0cm 40cm 0cm}, clip, width=\linewidth]{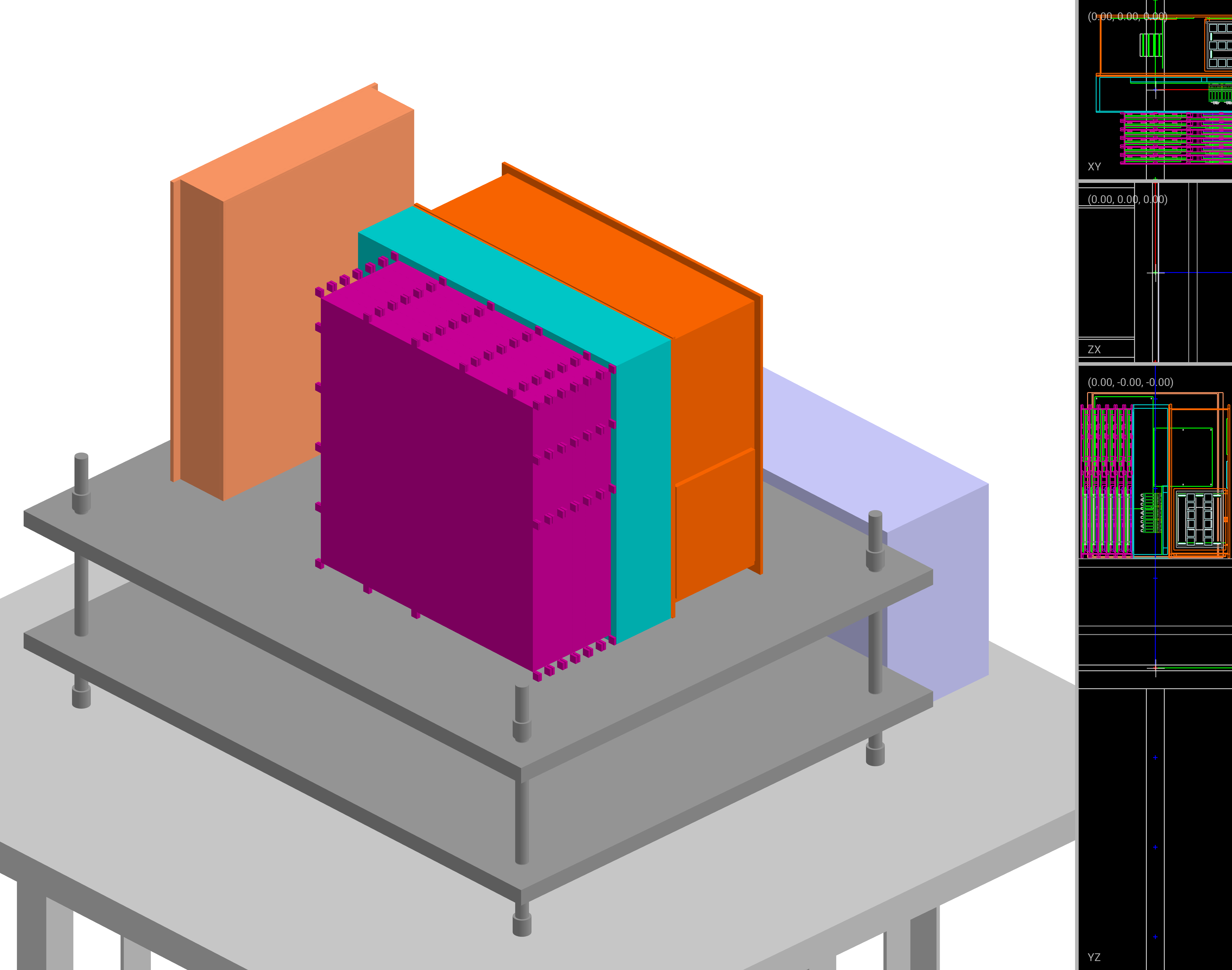}
  \caption{CAD model of the SWORD simulation modeling the ComPair system and environment experienced in the HIGS beam test.}
  \label{fig:sword}
\end{figure}

\begin{figure}[h!]
  \centering
  \includegraphics[trim={0cm 0cm 0cm 1.4cm}, clip, width=\linewidth]{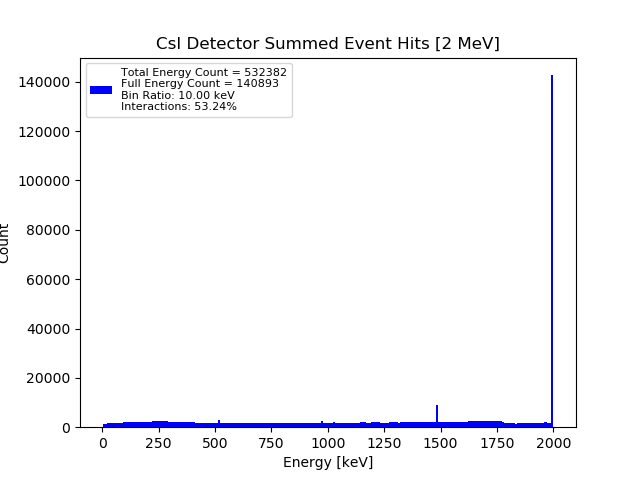}
  \caption{Simulated response of the CsI Calorimeter to 2 MeV gamma rays.}
  \label{fig:2MeV}
\end{figure}

\begin{figure}[h!]
  \centering
  \includegraphics[trim={0cm 0cm 0cm 1.4cm}, clip, width=\linewidth]{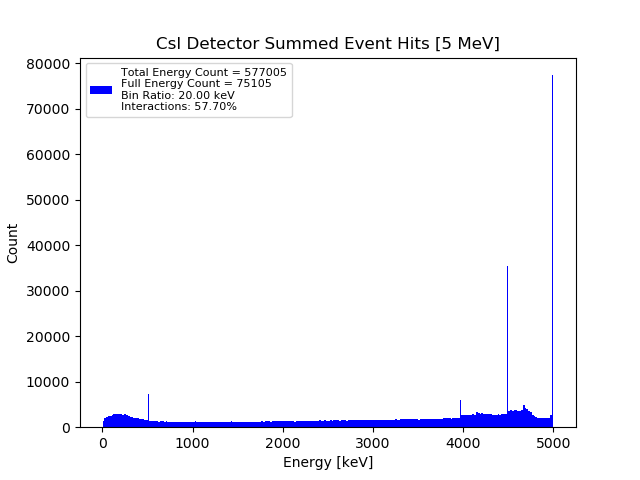}
  \caption{Simulated response of the CsI Calorimeter to 5 MeV gamma rays.}
  \label{fig:5MeVSim}
\end{figure}

\begin{figure}[h!]
  \centering
  \includegraphics[trim={0cm 0cm 0cm 1.4cm}, clip, width=\linewidth]{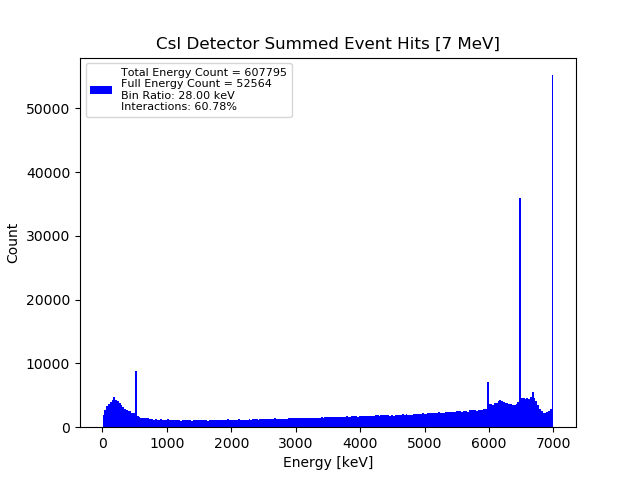}
  \caption{Simulated response of the CsI Calorimeter to 7 MeV gamma rays.}
  \label{fig:7MeV}
\end{figure}

\begin{figure}[h!]
  \centering
  \includegraphics[trim={0cm 0cm 0cm 1.4cm}, clip, width=\linewidth]{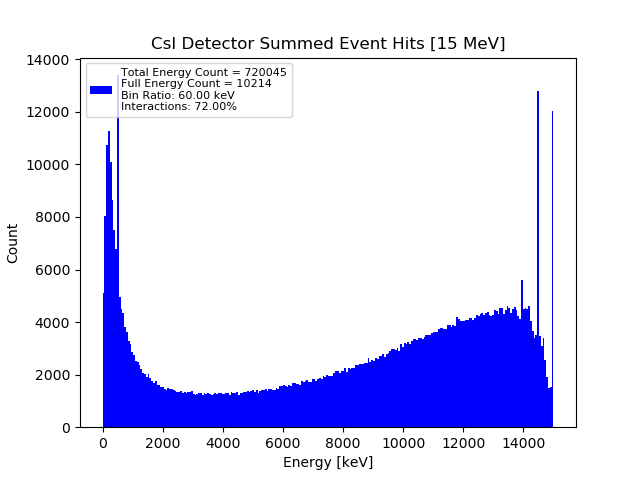}
  \caption{Simulated response of the CsI Calorimeter to 15 MeV gamma rays.}
  \label{fig:15MeV}
\end{figure}

\begin{figure}[h!]
  \centering
  \includegraphics[trim={0cm 0cm 0cm 1.2cm}, clip, width=\linewidth]{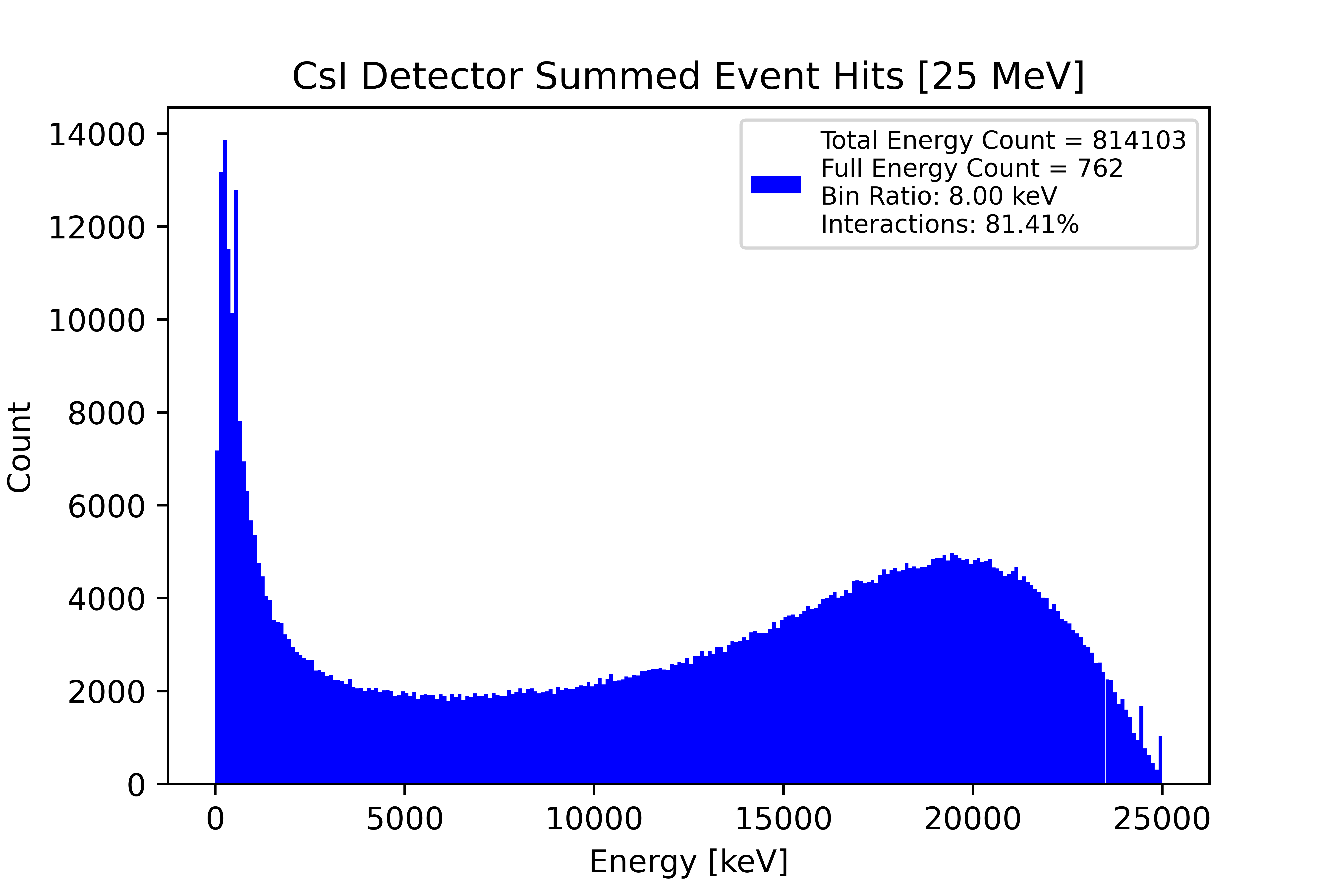}
  \caption{Simulated response of the CsI Calorimeter to 25 MeV gamma rays.}
  \label{fig:25MeV}
\end{figure}

This section presents simulations of the experimental setup as the HIGS facility using the SoftWare for Optimization of Radiation Detectors (SWORD) simulation package~\cite{SWORD}. Like the experimental setup, we shot a $0.5 \ \mathrm{mm}$ pencil beam into the system. Fig.~\ref{fig:sword} shows the system modeled in SWORD. Figs.~\ref{fig:2MeV},~\ref{fig:5MeVSim},~\ref{fig:7MeV},~\ref{fig:15MeV}, and~\ref{fig:25MeV} plots the full summed response of the calorimeter to 2, 5, 7, 15, and 25 MeV beams. Note that in the simulations, we did not add any energy blurring due to calorimeter performance as a means to display the underlying physical process. In the plots, we can see that the photopeak to continuum ratio decreases with energy and the escape peaks become more prominent with increasing energy. The experimental $5 \ \mathrm{MeV}$ full-energy peak appears less prominent than in simulation is likely due to the experienced broadening of the escape peaks and sitting on top of the Compton/pair continuum. Next, at higher energy, we can see the Compton edge feature disappear and become more of a continuum distribution. This is a result of the decrease in full-containment and leakage of the electromagnetic showers.

\bibliographystyle{IEEEtran.bst}
\bibliography{IEEEbib}

\end{document}